\documentclass[aps,twocolumn,showpacs]{revtex4}
\pdfoutput=1 
\usepackage{graphicx}
\usepackage{amsmath}

\newcommand{\rem}[1]{}

\begin{document}

\title{Internal and External Resonances of Dielectric Disks}

\author{C. P. Dettmann$^{1}$, G. V. Morozov$^{1}$, M. Sieber$^{1}$, H. Waalkens$^{1,2}$}
\affiliation{$^{1}$Department of Mathematics, University of Bristol, Bristol BS8 1TW,
United Kingdom\\
$^{2}$Department of Mathematics, University of Groningen, 9747 AG Groningen, The Netherlands}

\date{\today}

\begin{abstract}
Circular microresonators (microdisks)
are micron sized dielectric disks embedded in a material of lower refractive index.
They possess modes with complex eigenvalues (resonances)  which are solutions 
of analytically given transcendental equations.
The behavior of such eigenvalues in the small opening limit, 
i.e. when the refractive index of the cavity goes to infinity, is analysed. 
This analysis allows one to clearly distinguish between internal (Feshbach) 
and external (shape) resonant modes for both TM and TE polarizations. 
This is especially important for TE polarization 
for which internal and external resonances can be found in the same region 
of the complex wavenumber plane.  
It is also shown that for both polarizations, 
the internal as well as external resonances can be classified 
by well defined azimuthal and radial modal indices. 
\end{abstract}


\maketitle


\section{Introduction} 

Thin dielectric microcavities of various shapes
filled with a homogeneous material
are key components for the construction of optical microresonators and microlasers 
\cite{Vahala2003,Nosich2007}.
Their eigenmodes (resonances) are characterized 
by complex wavenumbers $k=k_r + i k_i$,
which are complicated solutions of 3D Maxwell equations.
However, the modes of microcavities, 
with the thickness only a small fraction of the mode wavelength, 
can be studied in 2D formulation 
with the aid of an effective refractive index $n_{\rm eff}$
which takes into account the material as well as the thickness
of the cavity, see for example appendix~I of Ref.~\cite{Nosich2005},
or chapter~II of Ref.~\cite{Bogomolny2007}. 
Among such microcavities, 
circular cavities are one of very few cases 
where the transcendental equations for  
complex eigenmodes (resonances) can be found analytically.
The analysis of these equations shows that
there exist two kinds of resonances, see for example Ref.~\cite{Bogomolny2008}.
Following common terminology we will call them 
internal (or Feshbach) and external (or shape) resonances.
In general, external  resonances have relatively large imaginary parts compared to internal resonances.
The external resonances are thus very leaky 
(i.e. have low $Q$-factors defined as $Q = {{k_r } \mathord{\left/{\vphantom {{k_r } {2k_i }}} \right.
\kern-\nulldelimiterspace} {2k_i }}$) and, as a result, 
cannot be directly used for lasing.
But occasionally they can occur in the same wavenumber range as the internal resonances
and therefore cannot always be ignored.
This deserves systematic investigation.

The purpose of this letter is twofold.
First, the behaviour of the circular cavity (disk) resonances
in the small opening limit, 
i.e. when the refractive index of the cavity diverges, is analysed. 
We note that for internal resonances this has recently been studied in Ref.~\cite{Lee2008}. 
For completeness we reproduce their results, 
though using a mathematically different and more illustrative approach. 
Our analysis allows us to clearly distinguish between internal 
and external resonant modes for 
both transverse magnetic (TM; electric field perpendicular to the disk plane)
and transverse electric (TE; magnetic field perpendicular to the disk plane) 
polarizations of the electromagnetic field. 
This is especially important for TE polarization 
for which internal and external resonances 
can be found in the same region of the complex wavenumber plane.  
Second, it is shown with the aid of the above limit,
that both internal and external resonances can be classified 
by well defined azimuthal and radial modal indices for both polarizations. 


\section{Equations for Resonances} 

Let $\Psi$ stand for $E_z$ in the case of TM polarization
and for $H_z$
in the case of TE polarization,
where $E_z$ and $H_z$ are electric and magnetic fields respectively.
For a homogeneous dielectric microdisk of radius $R$
and effective refractive index $n$ in a medium
of refractive index $1$, Maxwell's equations reduce to
\begin{equation}
\label{eq:Psi}
 \frac{{\partial ^2 \Psi }}{{\partial r ^2 }} + \frac{1}{r }\frac{{\partial \Psi }}{{\partial r }} + \frac{1}{{r ^2 }}\frac{{\partial ^2 \Psi}}{{\partial \varphi ^2 }} + k^2 n^2 \Psi \left( {r ,\varphi } \right) = 0,
\end{equation}
inside the microdisk ($r<R$) and the same form with $n$ replaced by $1$ 
outside the microdisk ($r>R$).
The resonances  are obtained by imposing outgoing boundary conditions
at infinity, i.e. we require that 
$\Psi \left( r \right) \propto {{e^{ikr} } \mathord{\left/
 {\vphantom {{e^{ikr} } {\sqrt r }}} \right.
 \kern-\nulldelimiterspace} {\sqrt r }}$,
$r \rightarrow \infty$.
For physical reasons, the value of the EM field at the disk center 
must be finite.
These boundary conditions in combination with the continuity 
of the electric field $E_z$ and its derivative for TM modes 
(or the magnetic field $H_z$ and its derivative divided by the square of the refractive index for TE modes)
at $r=R$ lead to the resonant field $\Psi$ in the form of twofold degenerate (for $m>0$) 
whispering gallery (WG) modes
\begin{equation}
\label{eq:wave_function}
\Psi _z^m  = \left\{ {\begin{array}{*{20}c}
   {N_m J_m \left( {knr} \right)\left({\begin{array}{*{20}c}
   {\cos m\varphi }  \\
   {\sin m\varphi }  \\
\end{array}} \right), \quad r < R,}  \vspace{8pt}\\
   {H_m \left( {kr} \right)\left( {\begin{array}{*{20}c}
   {\cos m\varphi }  \\
   {\sin m\varphi }  \\
\end{array}} \right), \quad r > R,}  \\
\end{array}} \right.
\end{equation}
where for TM modes the complex wavenumbers $k$ are solutions of
\begin{equation} 
\label{eq:unpert_quant_cond_tm}
J_m( k n R )H_m^{\,\,\,'} (k R) 
- n\,J_m^{\,\,\,'} ( k n R )H_m ( k R ) = 0,
\end{equation}
and for TE modes the complex wavenumbers $k$ are solutions of
\begin{equation} 
\label{eq:unpert_quant_cond_te}
J_m( k n R )H_m^{\,\,\,'} (k R) 
-\frac{1}{n}J_m^{\,\,\,'} ( k n R )H_m ( k R ) = 0.
\end{equation}
Here $J_m$ and $H_m$ are Bessel and Hankel functions of the first kind respectively, 
$m=0, 1, 2, ...$ is the azimuthal modal index,
and $N_m=H_m(kR)/J_m(knR)$ are constants.
Physically, the azimuthal modal index $m$ characterizes 
the field variation along the disk circumference, with the number of intensity hotspots being equal to $2m$.
The radial modal index $q=1, 2, ...$ will be used to label different resonances
with the same azimuthal modal index $m$. 
We will discuss mathematical and physical interpretations
of the radial modal index $q$ in the next sections. 

In general, to study resonant modes we firstly solve numerically
Eq. (\ref{eq:unpert_quant_cond_tm}) and Eq. (\ref{eq:unpert_quant_cond_te})
for several azimuthal modal indices $m$ and for the fixed refractive index $n=1.5$,
using as initial guesses a fine grid in the complex wavenumber plane.
Then we numerically continue the solutions for increasing and decreasing $n$. 


\section{TM modes} 
 
Figures~1~and~2 show the behaviour of TM resonances given by the solutions of 
Eq.~(\ref{eq:unpert_quant_cond_tm}) for several azimuthal modal indices $m$ 
under variation of the refractive index $n$.
For a fixed $n$, like for $n=1.5$ in Figs.~1~and~2, one can clearly distinguish 
between the internal  and external  resonances
as they are located in well separated regions of the complex wavenumber plane:
the internal resonances have much smaller imaginary parts in comparison  to the external  resonances.
For each of the two kinds of resonances with the same $m$,
we consecutively assign a radial modal index $q$ 
in accordance with the increase of their real parts $k_r$ starting from $q=1$.

\begin{figure}[htb]
\includegraphics[width=8.4cm]{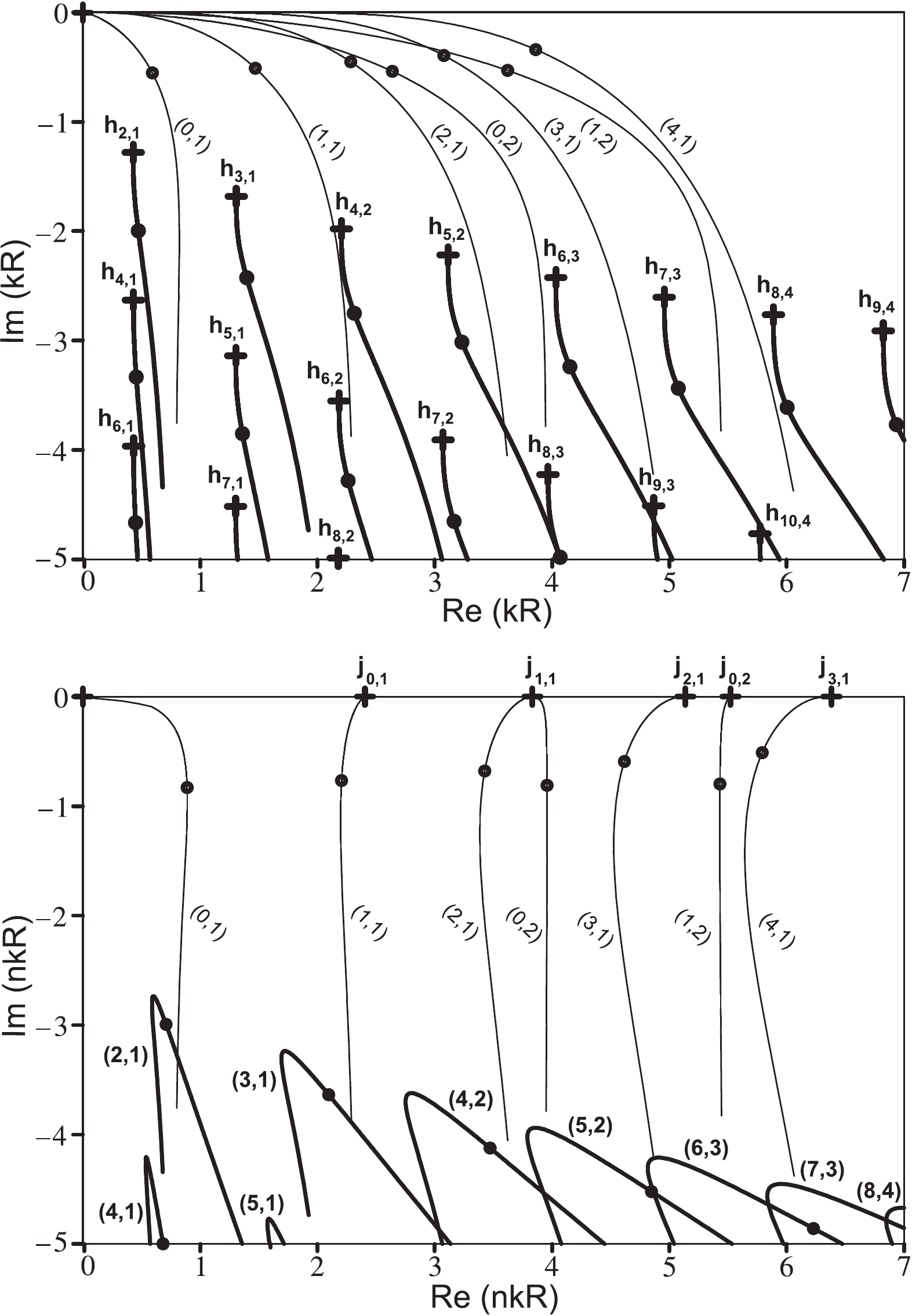}
\caption{TM internal (thin curves) and external (thick curves) resonances
of a dielectric microdisk of radius $R$ and refractive index $n$
varying from $n=1.001$ (loose ends) to infinity (crosses)
in the complex $kR$ plane (upper panel) and $nkR$ plane (lower panel). 
The filled circles correspond to $n=1.5$.} 
\label{Fig1}
\end{figure}

\begin{figure}[htb]
\includegraphics[width=8.4cm]{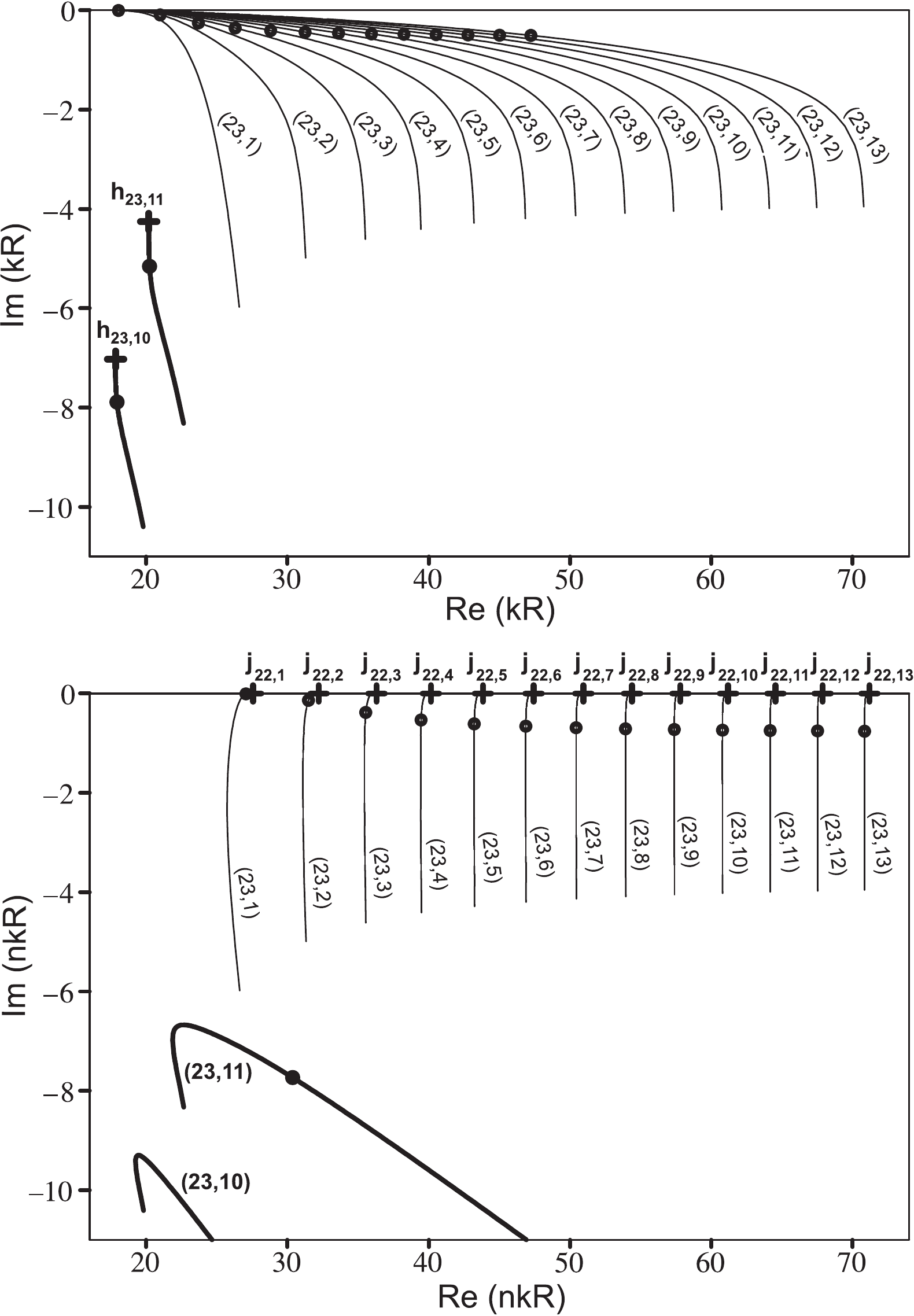}
\caption{TM internal (thin curves) and external (thick curves) resonances
with the azimuthal modal index $m=23$ 
of a dielectric microdisk of radius $R$ and refractive index $n$
varying from $n=1.001$ (loose ends) to infinity (crosses)
in the regions $16< {\rm Re}(kR)<74$, $-11<{\rm Im}(kR)<0$ 
of the complex $kR$ plane (upper panel)
and $nkR$ plane (lower panel). 
The filled circles correspond to $n=1.5$.} 
\label{Fig2}
\end{figure} 

Another difference between internal and external resonances 
(in addition to their location in the complex wavenumber plane) 
is the number of radial modes in each group of fixed azimuthal index $m$.
While there are infinitely many internal resonances for each azimuthal index $m\ge 0$, 
there are, as we will show below, only a finite number of external resonances for a given $m$, 
namely, none if $m$ is 0 or 1,   
$\frac{m}{2}$ if $m$ is even, and $\frac{m-1}{2}$ if $m$ is odd.

For TM internal resonances, the physical meaning of the radial modal index $q$ is
the number of intensity hotspots in the radial direction inside of the disk, see Fig.~3.
For TM external resonances, the index $q$ has no similar physical interpretation.
These resonances are so deep in the complex wavenumber plane that
the corresponding Bessel functions, see Eq.~(\ref{eq:wave_function}),
have almost no variation inside the disk.
This is illustrated in Fig.~4.

\begin{figure}[htb]
\includegraphics[width=8.4cm]{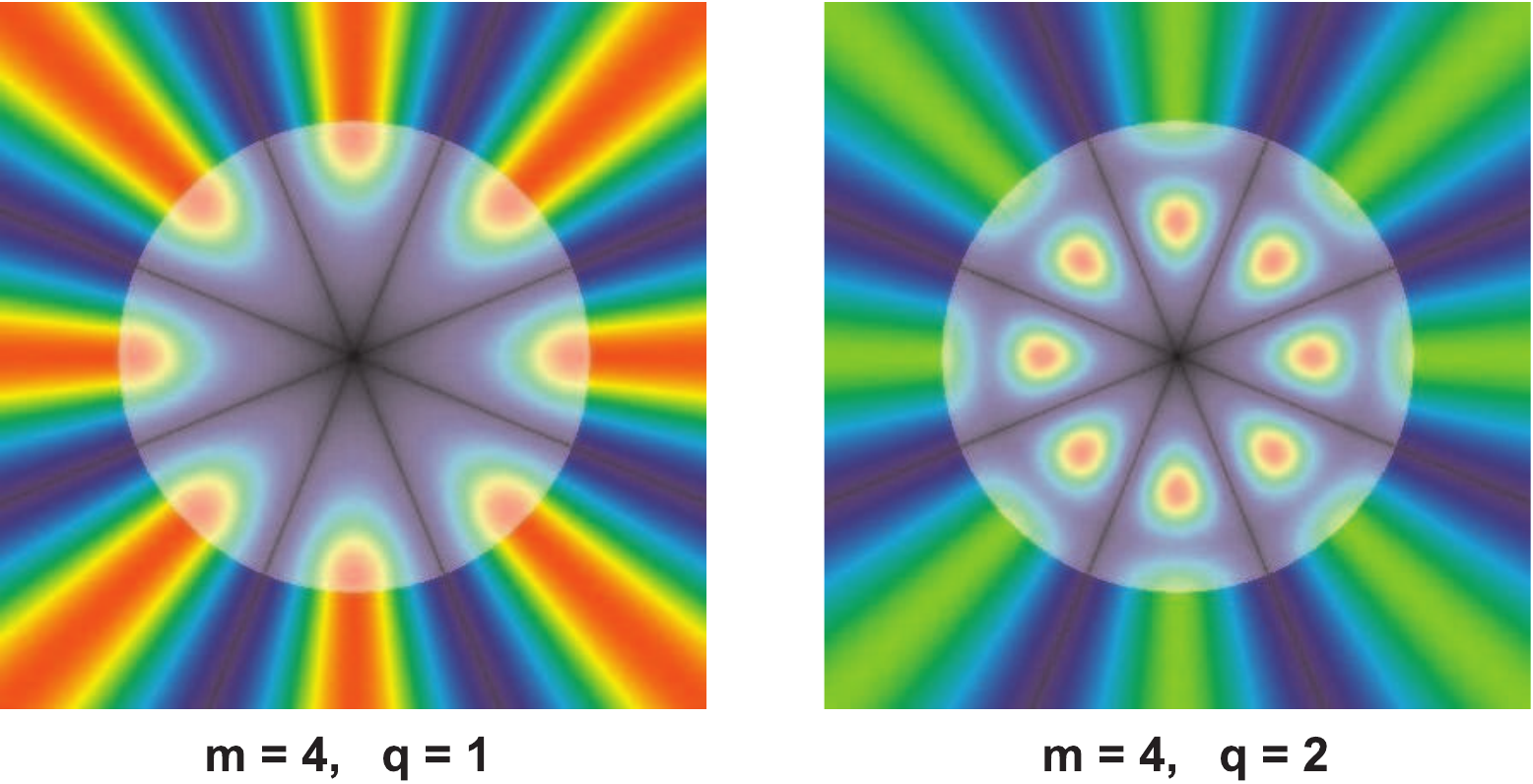}
\caption{(Color online). The intensity of TM internal resonant modes with the indicated modal indices 
in near-field region of the dielectric disk with $n=1.5$, $R=1$. 
Red indicates high intensities, purple/black indicates low  intensities.} 
\label{Fig3}
\end{figure} 

\begin{figure}[htb]
\includegraphics[width=8.4cm]{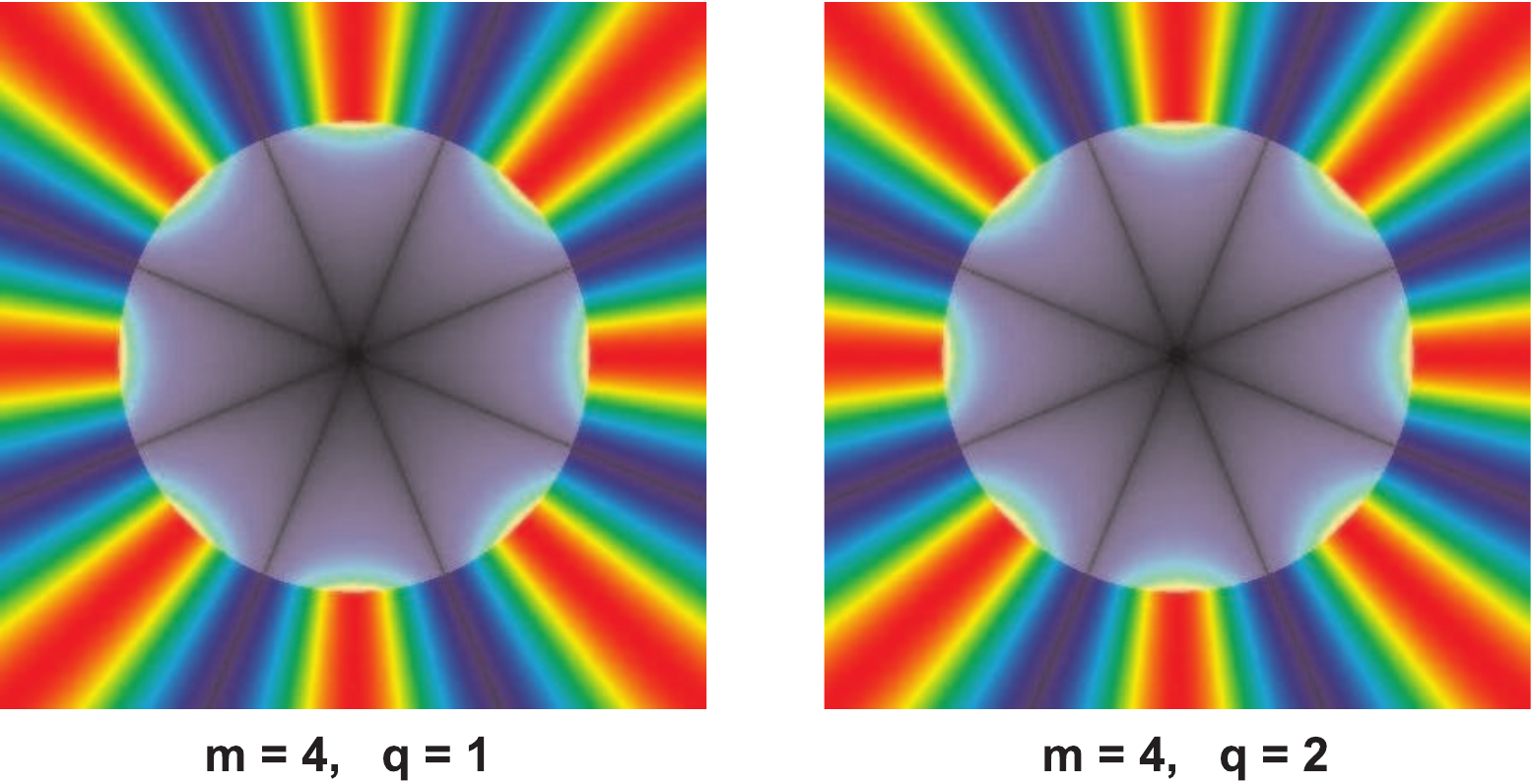}
\caption{(Color online). The intensity of TM external resonant modes with the indicated modal indices 
in near-field region of the dielectric disk with $n=1.5$, $R=1$.} 
\label{Fig4}
\end{figure} 

Let us now study the behaviour of the scaled wavenumbers  $nk_{m,q}R$
of TM internal resonances in the small opening limit, $n\rightarrow \infty$.
One would intuitively expect that in such a limit these resonances 
reduce to the real eigenvalues of a closed disk which are given 
by the corresponding zeros of Bessel functions $j_{m,q}$.
However, as we will see below, this is not the case.
In fact, for the TM internal resonances, see the thin curves in Figs.~1~and~2, we have
\begin{equation}
\begin{split}
   \mathop {\lim }\limits_{n \to \infty } nk_{m,q}R & = j_{m - 1,q}, \quad m \ne 0,  \\
   \mathop {\lim }\limits_{n \to \infty } nk_{0,q}R & = j_{1,q - 1}, \quad \,\,q \ne 1,  \\
   \mathop {\lim }\limits_{n \to \infty } nk_{0,1}R & = 0.
\end{split}
\end{equation}
This can be obtained by rewriting Eq.~(\ref {eq:unpert_quant_cond_tm}) as 
\begin{equation}
\frac{1}{n}\frac{{J_m (knR)}}{{J_{m + 1} (knR)}} = \frac{{H_m \left( {kR} \right)}}{{H_{m + 1} \left( {kR} \right)}}.
\nonumber
\end{equation}
Then, from inspecting  Figs.~1~and~2 we see that the quantity $nkR$ converges 
to finite real values and the quantity $kR$ converges to $0$ 
as $n\rightarrow \infty$ for all TM internal resonances.
But for  $kR \rightarrow 0$ we have
\begin{equation}
\frac{{H_m \left( {kR} \right)}}{{H_{m + 1} \left( {kR} \right)}}  \sim \left\{ {\begin{array}{*{20}c}
   {kR/(2m)\,,\quad \quad \quad \quad \quad \quad \quad \quad m > 0,}  \\
   {\left(i\pi/2 - \ln(kR/2) - \gamma \right)kR\,, \quad m = 0,}  \\
\end{array}} \right.
\nonumber
\end{equation}
where $\gamma=0.5772 \ldots$ is the Euler-Mascheroni constant.
As a result, we obtain for $n \rightarrow \infty$ and $m \neq 0$ 
\begin{equation}
\frac{1}{n}  \frac{J_{m}(k n R)}{J_{m+1}(k n R)} \rightarrow \,\frac{kR}{2m}  \nonumber\,,
\end{equation}
or equivalently 
\begin{equation}
J_{m-1}( k n R ) =  \frac{2m}{knR}  J_{m}(k n R) - J_{m+1}(k n R)  \rightarrow 0, \nonumber
\end{equation}
i.e. all $n$ scaled TM resonances wavenumbers $nk_{m \neq 0,q}R$ approach the zeros $j_{\,m-1,\,q}$ (rather than $j_{\,m,\,q}$).
Then, for $n \rightarrow \infty$ and $m=0$ we have
\begin{equation}
\frac{knR\,J_{1}(k n R)}{J_{0}(k n R)} \sim \frac{1}{i\pi/2 - \ln(kR/2) - \gamma}    \to 0    \nonumber\,.
\end{equation}
Since $J_0$, $J_1$ are regular along the real axis, 
we have that $nk_{0,1}R \to 0$ and $nk_{0,q \neq 1}R \to j_{1,q-1}$.

As for TM external resonances, they are all deep in the complex wavenumber plane.
But the zeros of $J_m$ are all real for $m \geq 0$.
Therefore, for the TM external resonances we have
\begin{equation}
\frac{1}{n}  \frac{J_{m}(k n R)}{J_{m+1}(k n R)} \rightarrow \,0\,, 
\quad n \rightarrow \infty  \nonumber\,.
\end{equation}
This immediately leads to
\begin{equation}
H_m(kR) \rightarrow \,0\,, 
\quad n \rightarrow \infty  \nonumber\,,
\end{equation}
i.e. all TM external resonance wavenumbers (not scaled with respect to $n$)
satisfy the relation
\begin{equation}
\mathop {\lim }\limits_{n \to \infty } k_{m,q}R  = h_{m,q},
\end{equation}
where $h_{m,q}$ are complex zeros of Hankel functions.
It is known, see Ref.~\cite{Erdelyi1953}, 
that there is only a finite number of such zeros for a given $m$:
$0$ if $m$ is 1 or 2, $m/2$ if $m$ is even, and $(m-1)/2$ if $m$ is odd.
This exactly corresponds to our findings for the number of radial modes
in each group of external resonances with fixed $m$.


\section{TE modes}

Figures~5~and~6 show the behaviour of TE resonances given by the solutions of 
Eq.~(\ref{eq:unpert_quant_cond_te}).
Like for TM polarization, we separate internal and external resonances with the same $m$
and assign the radial modal index $q$ to each member of those two sets independently,
in accordance with the increase of their real parts $k_r$ starting from $q=1$.
But there could be a problem. 
The TE external resonances 
$k_{m,\frac{m}{2}}$ with  $m=2,4,\ldots$
and $k_{m,\frac{m+1}{2}}$ with $m=1,3,\ldots$,
i.e. the last ones in the sets of fixed $m$,
do not necessarily have large imaginary parts.
(We will call these resonances the ``special" ones.)
As a result, they could be mixed with TE internal resonances
and their field intensities could display
some features of internal resonances as well.  
Therefore, the only way to separate them is 
to trace them with increasing $n$
till they reach their limits (when $n \rightarrow \infty$), 
see the thick and thin curves in Figs.~5~and~6.

\begin{figure}[htb]
\includegraphics[width=8.4cm]{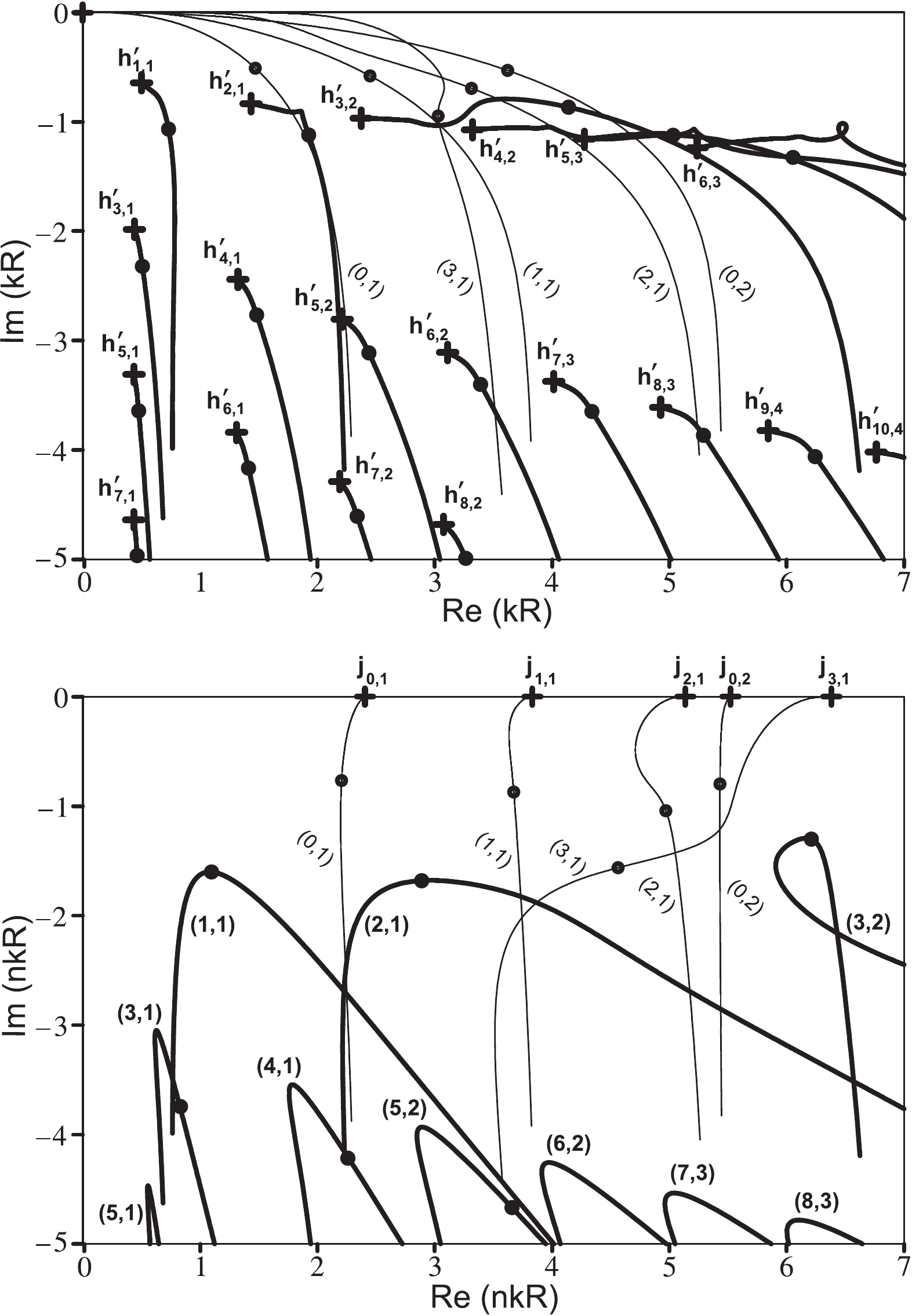}
\caption{
TE internal (thin curves) and external (thick curves) resonances
of a dielectric microdisk of radius $R$ and refractive index $n$
varying from $n=1.001$ (loose ends) to infinity (crosses)
in the complex $kR$ plane (upper panel) and $nkR$ plane (lower panel). 
The filled circles correspond to $n=1.5$. } 
\label{Fig5}
\end{figure} 

\begin{figure}[htb]
\includegraphics[width=8.4cm]{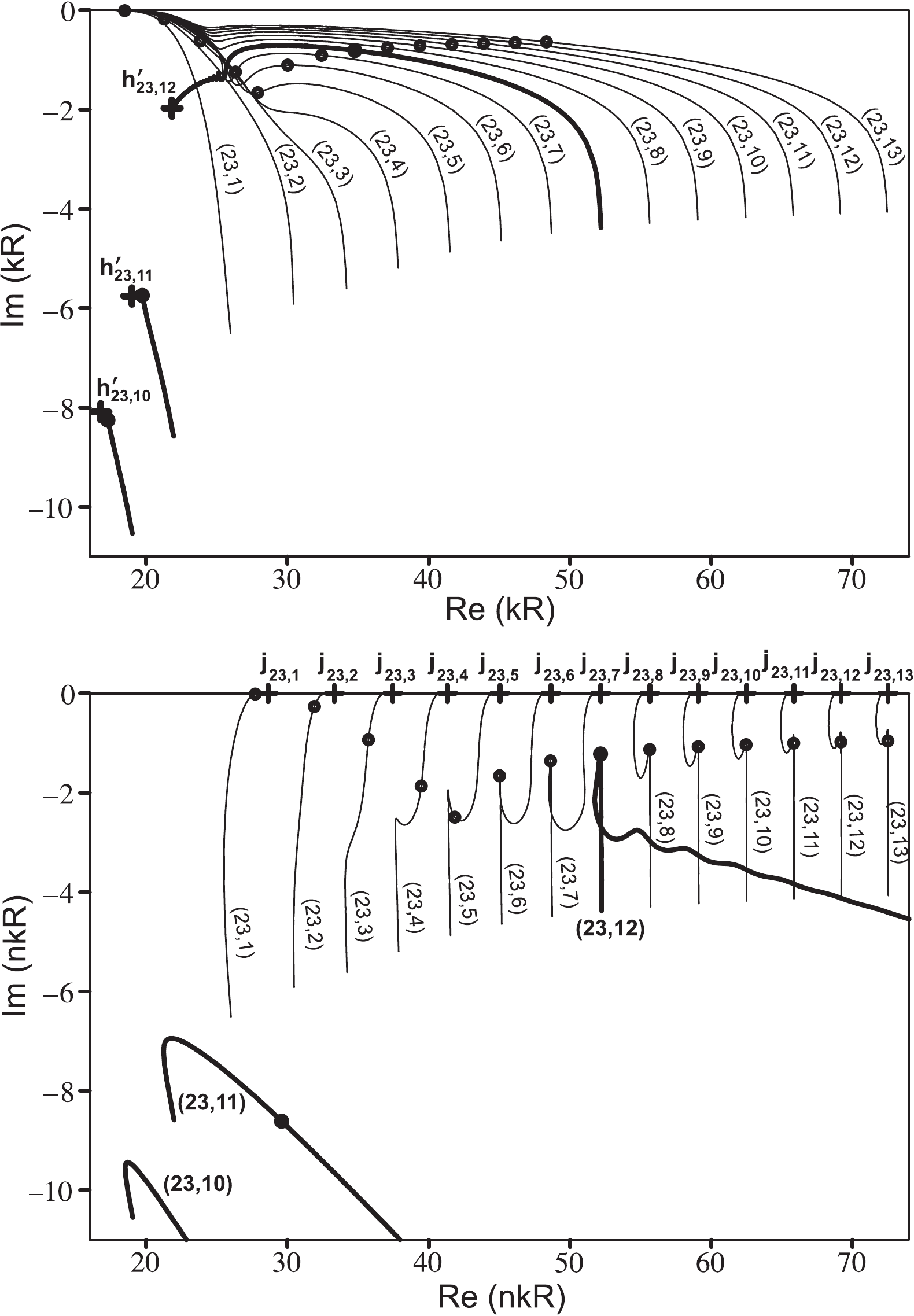}
\caption{
TE internal (thin curves) and external (thick curves) resonances
with the azimuthal modal index $m=23$ 
of a dielectric microdisk of radius $R$ and refractive index $n$
varying from $n=1.001$ (loose ends) to infinity (crosses)
in the regions $16 < {\rm Re}(kR) < 74$, $-11 < {\rm Im}(kR) < 0$ 
of the complex $kR$ plane (upper panel) and $nkR$ (lower panel). 
The filled circles correspond to $n=1.5$.} 
\label{Fig6}
\end{figure} 

Using arguments similar to the ones in the previous section on TM modes we find that 
Eq. (\ref{eq:unpert_quant_cond_te}) takes the form
$J_{m}( k n R ) = 0$ for internal and $H_{m}^{\,\,'}( k  R ) = 0$ for external resonances when $n \rightarrow \infty$.
This means that for the scaled wavenumbers of the TE internal resonances 
\begin{equation}
\mathop {\lim }\limits_{n \to \infty } nk_{m,q}R  = j_{m,q}. 
\end{equation}
as we intuitively expected.
The TE external resonances (not scaled with respect to $n$)
approach the complex zeros $h^{\,'}_{m,q}$ of
the corresponding Hankel function derivatives 
\begin{equation}
\mathop {\lim }\limits_{n \to \infty } k_{m,q}R  = h^{\,'}_{m,q}. 
\end{equation}
The thin and thick curves in Figs.~5~and~6 illustrate the results numerically.

\begin{figure}[htb]
\includegraphics[width=8.4cm]{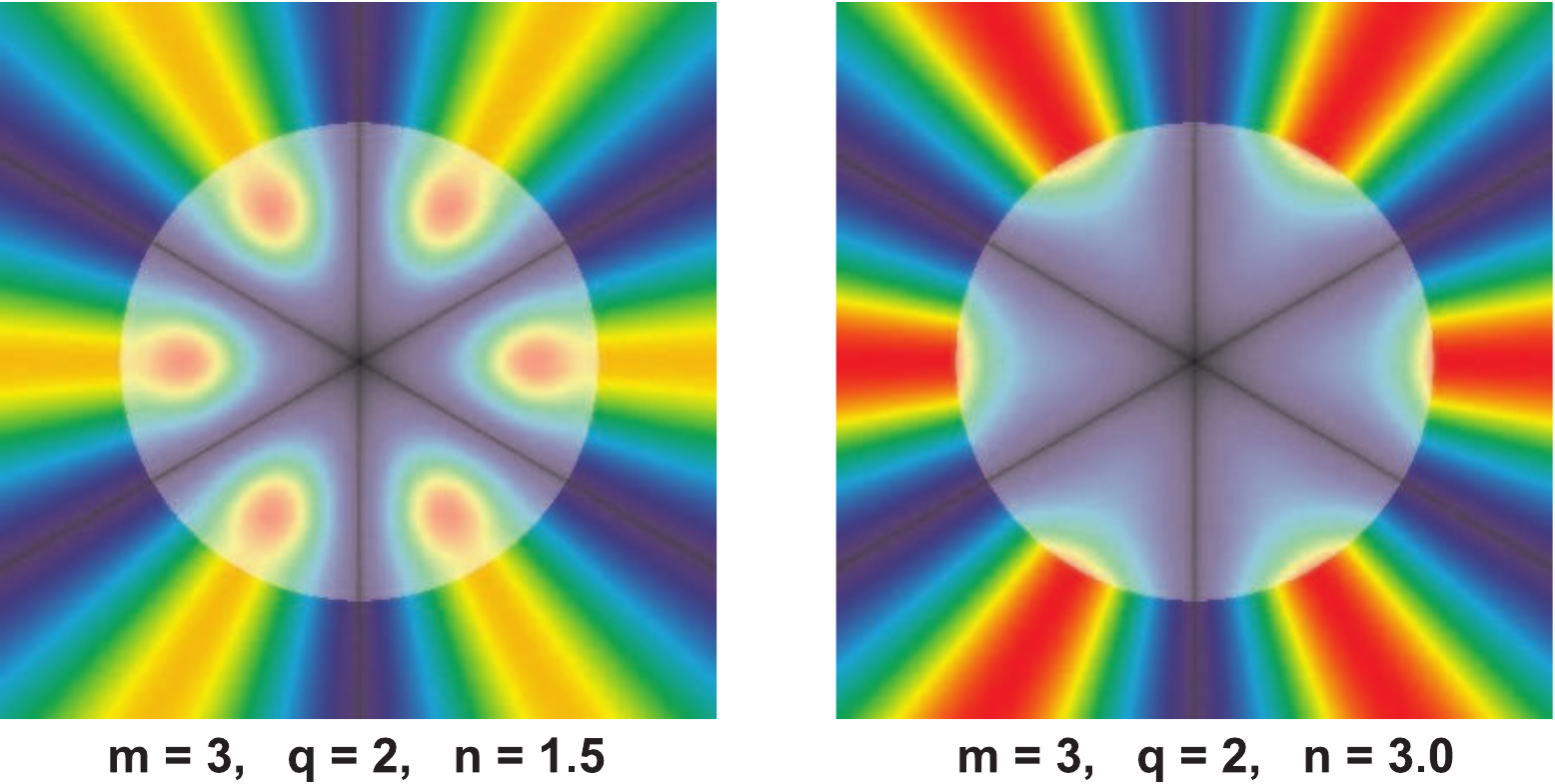}
\caption{(Color online). The intensity of TE external resonant modes with the indicated modal indices 
in near-field region of the dielectric disk with $R=1$ and $n=1.5$ (left panel),
$n=3.0$ (right panel).} 
\label{Fig7}
\end{figure} 

\begin{figure}[htb]
\includegraphics[width=8.4cm]{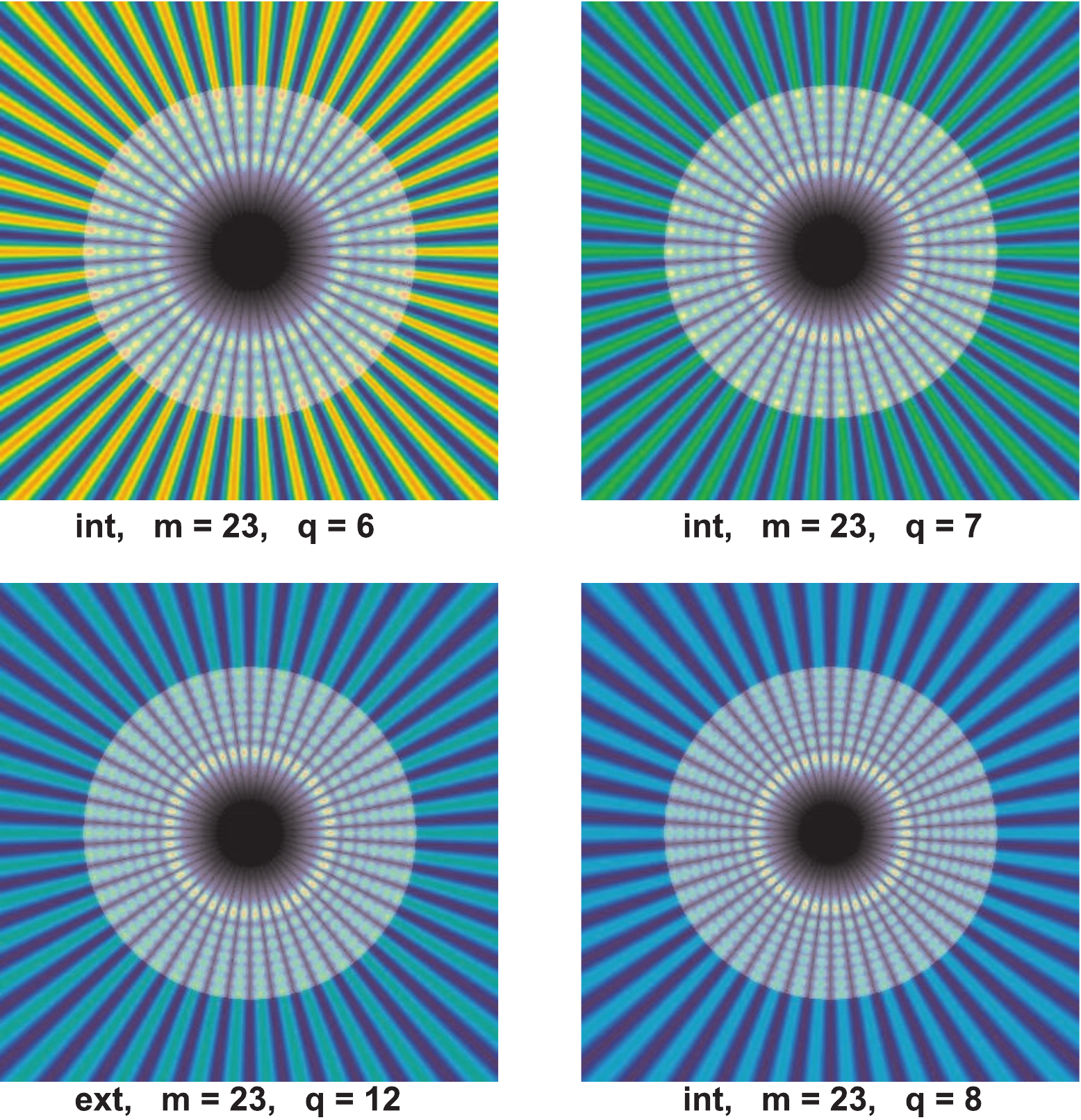}
\caption{(Color online). The intensity of TE resonant modes with the indicated modal indices 
in near-field region of the dielectric disk with $R=1$ and $n=1.5$; 
``int" stands for internal resonance, ``ext" stands for external resonance.} 
\label{Fig8}
\end{figure} 
The field intensities of most TE modes display pattens similar to TM modes:
the modal index $q$ for internal resonances gives the number of intensity hotspots in the radial direction;
for external resonances, which are deep in the complex wavenumber plane,
there is almost no field variation inside the disk.
However, the situation is different for the ``special'' TE external resonances. 
With the variation of the disk refractive index their imaginary parts could become relatively small. 
Then their field intensity patterns become similar to those of internal resonances,
see the left panel in Fig.~7.
Moreover, for large azimuthal indices $m$ and relatively low refractive indices $n$
the ``special" external resonances occupy positions exactly
where one would expect the corresponding internal resonances, see the modes with $n=1.5$ in Fig.~6. 
As a result, the field patterns of internal resonances to the left from the ``special" external one
display some unexpected features as well, see Fig.~8. 
For example, the field intensities of internal resonances $k_{23,6}$ and $k_{23,7}$
have five and six (rather than six and seven) intensity peaks in the radial direction.

\section{Conclusions} 

To summarize, we have studied in detail the behaviour of both internal (Feshbach)
and external (shape) resonances of an open microdisk in the small opening limit, 
i.e. when the microdisk refractive index diverges, for both TM and TE polarizations. 
Contrary to naive expectations, the limit values of the open disk resonances
match the eigenvalues of the corresponding closed disk with the zero (Dirichlet) 
boundary conditions only for TE internal resonances.
Our analysis assigns unambiguous azimuthal and  radial modal indices 
to each internal and external resonant mode.
We showed that the latter index has a clear physical interpretation
only for internal resonances, with one qualification.
As the refractive index $n$ is decreased one observes the striking phenomenon 
that some special TE external resonances join the set of internal resonances
and share their features.  
Our results should be of general interest 
since to the best of our knowledge this is the first complete classification 
of all resonant modes in the well-known and the simplest open system - 
a dielectric microdisk. 


\end{document}